\documentclass[twocolumn,showpacs,preprintnumbers,aps,prd,longbibliography]{revtex4-1}

\usepackage{hyperref}
\hypersetup{
    backref =       true,
    pagebackref  =  true,
    colorlinks =    true,
    linkcolor =     [rgb]{0.0,0.0,0.8},
    anchorcolor =   [rgb]{0.0,0.0,0.8},
    citecolor =     [rgb]{0.0,0.0,0.8},
    filecolor =     [rgb]{0.0,0.0,0.8},
    urlcolor =      [rgb]{0.0,0.0,0.8},
    pdftitle=       {Title},
    pdfsubject=     {Title},
    pdfauthor=      {A. Uthor}
}
\usepackage{nccmath}
\usepackage[utf8]{inputenc}
\usepackage{amsmath}
\usepackage{graphicx}
\makeatletter
\usepackage{etoolbox} 
\usepackage{lipsum} 
\usepackage[capitalize]{cleveref}

\@ifundefined{textcolor}{}
{
 \definecolor{BLACK}{gray}{0}
 \definecolor{WHITE}{gray}{1}
 \definecolor{RED}{rgb}{1,0,0}
 \definecolor{GREEN}{rgb}{0,1,0}
 \definecolor{BLUE}{rgb}{0,0,1}
 \definecolor{CYAN}{cmyk}{1,0,0,0}
 \definecolor{MAGENTA}{cmyk}{0,1,0,0}
 \definecolor{YELLOW}{cmyk}{0,0,1,0}
}

\usepackage{color, soul}

\usepackage{txfonts}
\usepackage{dcolumn}
\usepackage{bm}
\usepackage{natbib}
\usepackage{siunitx}
\usepackage{microtype}
\makeatother
\begin{document}

\title{Observation of Nonlinear Dynamics in an Optical Levitation System}

\author{Jinyong Ma, Jiayi Qin, Geoff T.\ Campbell, Giovanni Guccione, Ruvi Lecamwasam,  Ben C.\ Buchler, and Ping Koy Lam$^{}$}\email{Ping.Lam@anu.edu.au}

\affiliation{Centre for Quantum Computation and Communication Technology, Department of Quantum Science, Research School of Physics and Engineering, The Australian National University, Canberra ACT 2601, Australia}

\date{\today}
\begin{abstract}
\noindent\textbf{ABSTRACT} Optical levitation of mechanical oscillators has been suggested as a promising way to decouple the environmental noise and increase the mechanical quality factor. Here, we investigate the dynamics of a free-standing mirror acting as the top reflector of a vertical optical cavity, designed as a testbed for a tripod cavity optical levitation setup. To reach the regime of levitation for a milligram-scale mirror, the optical intensity of the intracavity optical field approaches \SI{3}{MW.cm^{-2}}. We identify three distinct optomechanical effects: excitation of acoustic vibrations, expansion due to photothermal absorption, and partial lift-off of the mirror due to radiation pressure force. These effects are intercoupled via the intracavity optical field and induce complex system dynamics inclusive of high-order sideband generation, optical bistability, parametric amplification, and the optical spring effect. We modify the response of the mirror with active feedback control to improve the overall stability of the system.
\end{abstract}


\maketitle

\section*{Introduction}
Cavity optomechanical systems, coupling light field with mechanical oscillators via radiation pressure force, have generated considerable interest in recent years, especially for systems in the quantum regime. They have found applications in precise metrology~\cite{AbbottObservation2016, KolkowitzCoherent2012, gavartin_force-measure_opto, krause_accelerometer_opto, QvarfortGravimetry2018, cervantes_accelerometer_opto, forstner_sensitivity_2012, forstner_magnetometer_opto}, nonreciprocal coupling~\cite{XuTopological2016, XuNonreciprocal2019}, preparation of mechanical quantum states~\cite{RiedingerNonclassicalcorrelationssingle2016}, entanglement between mechanical and optical systems~\cite{PalomakiEntanglingMechanicalMotion2013}, and may also be used to probe the fundamental physics of macroscopic quantum mechanics~\cite{Romero-IsartLargeQuantumSuperpositions2011, AlbrechtTestingquantumgravity2014} and test models of quantum gravity~\cite{PikovskiProbing2012a, YangMacroscopic2013}. A prerequisite to operate in the regime of quantum optomechanics is a high mechanical quality factor~\cite{marquardt_optomechanics_2009}, and thus a low mechanical decoherence rate, such that the coherent optomechanical coupling rate can be larger than both mechanical and optical decoherence rates\cite{VerhagenQuantumcoherent2012}. Some devices have indeed reached the motional ground state with the assistance of cryogenics and laser cooling~\cite{OConnellQuantumgroundstate2010, TeufelSideband2011, chan_cooling_opto}. Newer platforms are being designed to minimize the mechanical clamping of the resonator to exhibit high mechanical quality factor that can be used to explore quantum phenomena even at room temperature~\cite{NorteMechanicalResonatorsQuantum2016}.


The main channel by which environmental thermal noise enters the system is via mechanical supports. By forsaking any form of extrinsic clamping, mechanical oscillators can sustain coherent oscillations for extended times and would therefore pose as a better candidate for quantum optomechanics. Optical levitation \cite{SwartzlanderStableopticallift2010, marago_optical_2013, neukirch_multi-dimensional_2015, RahmanLaserrefrigerationalignment2017} has been identified as a promising route in this direction. In addition to the benefits of environmental isolation, these schemes allow us to fully manipulate the quantum state and mechanical frequency of the levitated mirror via the optical fields. In particular with the proposals of a scattering-free tripod~\cite{guccione_scattering-free_2013} or sandwich~\cite{MichimuraOptical2017} of optical cavities, a milligram-scale mirror is levitated by coherently interacting with the radiation pressure forces provided by optical resonators. Due to the scattering-free feature of these two schemes, the mechanical quality factor of the levitating mirror can be expected to be on the order of $10^{10}$ \cite{guccione_scattering-free_2013} and the quantum cooperativity can be estimated as $10^3$~\cite{MichimuraQuantum2020}, making them excellent candidates to explore quantum and nonlinear effects in the macroscopic regime.

Here we set up a reduced version of the optical tripod, where we consider only one vertical optical cavity as a testbed for levitation~\cite{LecamwasamDynamics2020}. This simplified setup enables us to investigate the system's dynamics and build a theoretical reference to better understand the underlying physics. Supported by a good agreement between experiment and numerical simulations, we believe that our models will be useful for any free-standing optomechanical system under high power. In particular, under the high power required for levitation, the optical intensity can reach \SI{3}{MW.cm^{-2}}, which is even larger than that of the Laser Interferometer Gravitational-wave Observatory (LIGO) \cite{AbbottObservation2016}. The response of our optomechanical system is dominated by different types of nonlinear interactions. The most prominent arises from the photothermal expansion of the mirror coating, which leads to optical bistability~\cite{AnOptical1997, HorvathPhotothermal2013} and a hysteretic asymmetry in the cavity response as a function of detuning. Another consequence of photo-absorption is the excitation of the acoustic modes of the mirror, which perturbs the cavity field. The fluctuation of the cavity field is subsequently propagated to other degrees of freedom via optical back action. In particular, we show that the absorbed energy results in effective anti-damping and parametric amplification~\cite{weng_stabilization_2015, BrachmannPhotothermal2016, AltinRobust2017}. Finally, the mirror also interacts with the intracavity field via the radiation pressure force, inducing a displacement when the optical push is stronger than the gravitational force. The interplay between all of these interactions results in complex behavior and rich dynamics. Higher-order optical sidebands~\cite{xiong_higher-order_2012, JiaoNonlinear2016, carmon_temporal_2005} and a quasi-continuous spectrum of the optical output are observed, which are suggestive of chaotic behaviour~\cite{marino_chaotically_2007, marino_chaotically_2011, marino_coexisting_2013}. To assess the stability of the cavity under external control, we implement active feedback~\cite{poggio_feedback_2007, CorbittOpticalDilutionFeedback2007} to suppress the excitation of the acoustic vibration. With the analysis and modeling drawn from these investigations, we show a route towards the realization of stable and coherent optical levitation.
\begin{figure*}[htb] \centering
\includegraphics[width=1\textwidth]{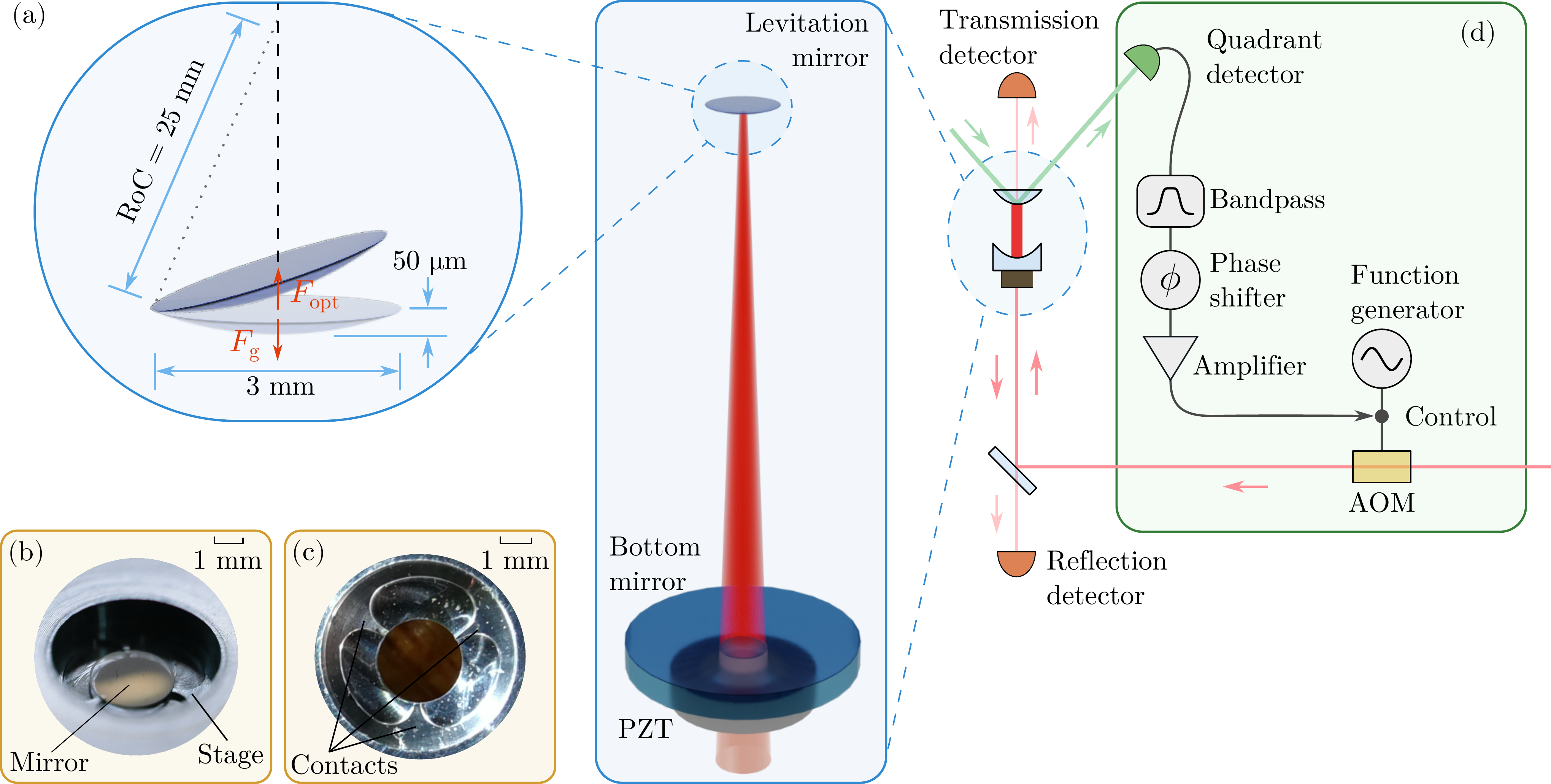} \caption{\textbf{Experimental setup.} (a) A small lightweight mirror at the top of a vertical optical cavity is subject to the radiation pressure force of the resonator field. The left pop-up shows the characteristics of the mirror. The mirror has a radius of curvature (RoC) of 25 mm. $F_{\rm opt}$ and $F_{\rm g}$ indicate the radiation pressure and gravitational forces respectively. The resonator is a vertical Fabry-P\'erot cavity with a piezoelectric actuator (PZT) on the bottom mirror. The input optical field is introduced at the bottom mirror. A small amount of the optical field reflected from the cavity bottom mirror transmits through the alignment mirror and is detected by a photodetector.~The cavity field transmitted through the levitation mirror is also measured. A low-intensity probe beam (green) introduced at an angle is reflected onto a quadrant detector to detect the displacement of the mirror.~(b) Photograph of the levitation mirror placed on the Invar mount. (c) Photograph of the supporting stage is carved directly out of the Invar cavity mount. The stage has three small contact points and a circular hole for optical access.~(d) The region enclosed within the green box outlines the feedback control, which will be discussed in the last section. This includes signal processing (bandpass filter, phase shifter, amplifier) and an acousto-optic modulator (AOM) to implement the control.}
\label{fig:1_NDL}
\end{figure*}
\section*{Results}
\noindent\textbf{Experimental setup.} Our experimental setup [Fig.~\ref{fig:1_NDL}(a)] consists of an optical resonator in a vertical configuration. The top mirror of the cavity, shown in Fig.~\ref{fig:1_NDL}(b) and hereon also referred to as the levitation mirror (or simply the mirror), is free-standing on a top of the Invar mount. The hole is specifically designed to have three symmetric contact points to minimize Van der Waals interactions [Fig.~\ref{fig:1_NDL}(c)]. When the radiation pressure force is sufficiently strong to balance the gravitational weight of the mirror, the torque exerted heaves a side off one of the contact points. As the mirror lifts, the effective cavity length increases, and as a consequence the laser driving the intracavity field becomes blue-detuned compared to the cavity's resonance. In this regime the optical spring effect provides a restoring force, which is the origin of the optical trap expected for the full tripod levitation setup~\cite{guccione_scattering-free_2013}.

\begin{figure*}[htb] \centering
\includegraphics[width=1\textwidth]{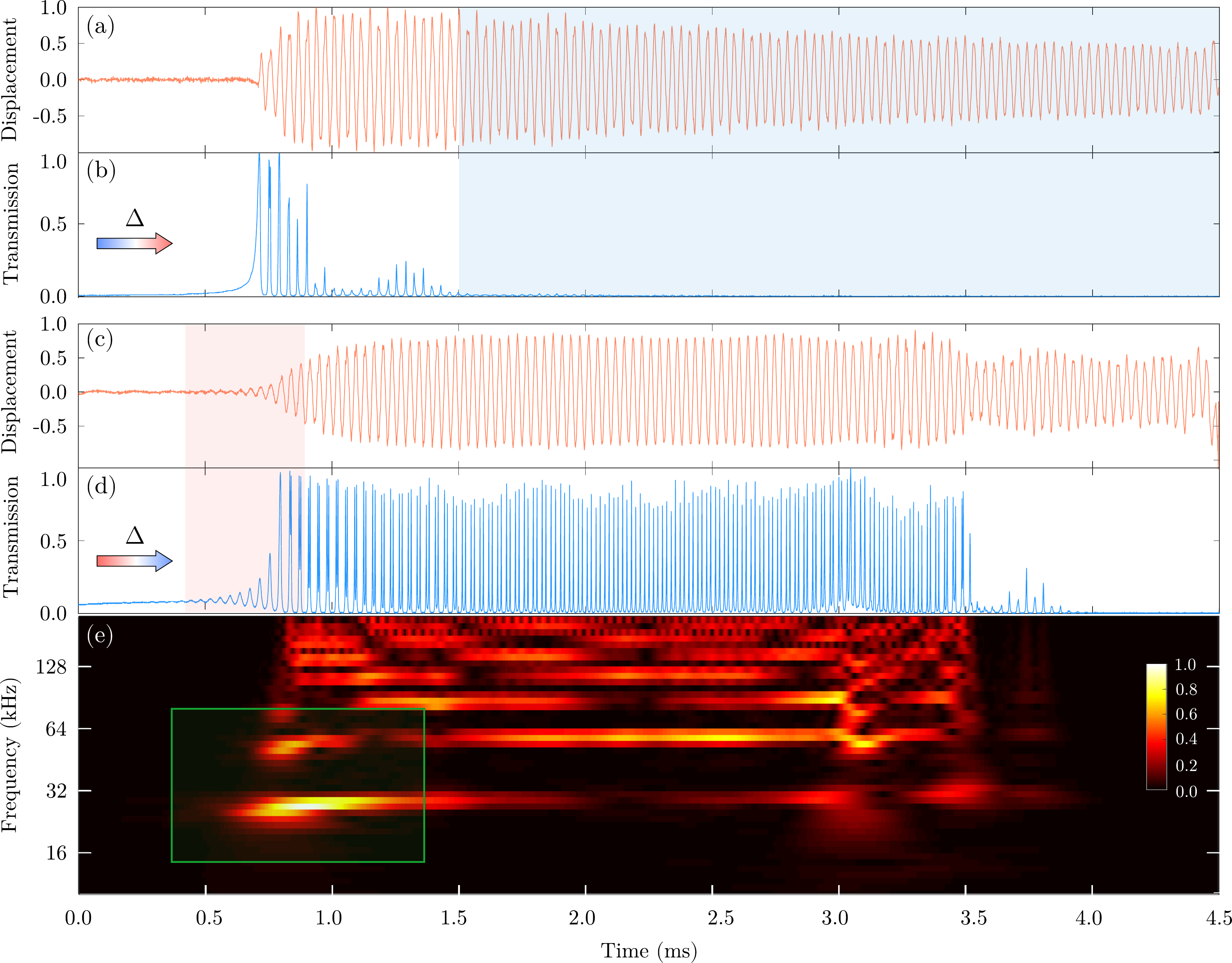} \caption{\textbf{System dynamics at an input power of \SI{4.5}{\watt}}. The estimated peak intracavity power is higher than the threshold of \SI{4}{\watt} to lift the mirror optically. Both cavity transmission and mechanical displacement are normalized to their maximum values respectively. The color change of the cavity detuning $\Delta$ indicates the scan direction.  (a)-(b) Displacement of the small mirror and corresponding cavity transmission during an anti-locking upward scan of the piezoelectric actuator (scanning from blue to red detunings). In the blue-shaded region the intracavity field vanishes while the oscillations perpetuate, allowing the opportunity to calibrate the mechanical frequency and damping rate of the excited mechanical mode. (c)-(d) Displacement and cavity transmission during a self-locking downward scan of the piezoelectric actuator (scanning from red to blue detunings). The cavity response is significantly distinct from the case of an upward scan. The red-shaded region indicates the onset of parametric amplification. (e) Wavelet transform of cavity transmission during self-locking. The green box highlights the shift in acoustic frequency as the average detuning is varied over time.}
\label{fig:4p5W}
\end{figure*}

To detect the mechanical displacement directly we introduce a weak optical beam at an angle that is reflected on the backside of the mirror and collected onto a quadrant photodetector. The intensity readout between different quadrants is then subtracted to obtain a relative measurement of the position. We also use the reflected and transmitted outputs of the main cavity field to monitor the evolution of the intracavity optical field. In Fig.~\ref{fig:4p5W} we show an example of the dynamics in the system to give an idea of the interactions involved. Note that all of the measurements presented in this and following figures are single time traces but are highly reproducible (see Supplementary Note 2 for details). We will unravel the different elements involved over the next few sections.

The most noticeable effect is the onset of mechanical oscillations (Fig.~\ref{fig:4p5W}) resulting in multiple crossings of the cavity resonance. When the average detuning is scanned linearly from higher to lower frequencies [Fig.~\ref{fig:4p5W}(a)-(b), piezoelectric actuator moving up shortening the length of the cavity], the resonance is quickly crossed, and the oscillation amplitude starts to decay overtime immediately after the brief excitation of the cavity field. Scanning in the opposite direction [Fig.~\ref{fig:4p5W}(c)-(d), actuator moving down extending the length of the cavity], one can instead observe a slow build-up of the oscillations even before the first full resonance crossing. Moreover, the oscillations become self-sustained, and the cavity enters a passive feedback loop that is broken only when the scan moves the average detuning too far. We label these two opposite responses as `anti-locking' and `self-locking' respectively, and we will see that they stem from a competition between photothermal and radiation pressure interactions \cite{marino_chaotically_2011, marino_coexisting_2013}. The observed oscillations are identified as excitations of the mirror's acoustic modes by the intracavity field. We also note two interesting aspects from the wavelet transform of the transmission output of the cavity during the self-locking scan in Fig.~\ref{fig:4p5W}(e). The first involves the generation of high-order sidebands of the acoustic modes induced by the nonlinearity of the system~\cite{CarmonDynamical2004, kippenberg_analysis_2005}. The second relates to the region enclosed by the green box, which shows how the intracavity field modifies the natural frequency of the acoustic mode. This is a consequence of the optical spring effect~\cite{SheardObservation2004, singh_all-optical_2010, MizrahiTwoslab2007, CorbittAllOptical2007, VogelOptically2003}.

\noindent\textbf{Equations of motion.} Before entering any in-detail analysis, it is important to develop a simple and effective model of the system to help understand the different phenomena. In our model we separate the position degree of freedom of the mirror into three different entities, each subject to a different type of interaction with the intracavity field: $x_{\rm th}$ accounts for displacements of the reflective coating due to photothermal expansion, $x_{\rm ac}$ corresponds to a different position of the mirror's surface following the vibrations of acoustic mode, and $x_{\rm lev}$ represents a full shift of the centre of mass because of radiation pressure force. All three interact with the optical degree of freedom, $a$, describing the amplitude of the optical field inside the cavity. Our model is far from being comprehensive: other phenomena may be present that are not accounted for, among which we mention for example, the bolometric interaction directly coupling photothermal absorption to the acoustic mode, the dependence of cavity decay rate on either acoustic or photothermal displacements, the Brownian motion and other stochastic dynamics of the three position degrees of freedom, the shot noise and phase fluctuations of the laser, or even the black body radiation of the levitating mirror. We will show, however, that a simpler model based simply on the mutual interaction of the cavity field with the three position degrees of freedom identified before is sufficient to reconstruct a faithful picture of the whole system, in good agreement with the experimental results.

We therefore consider the following equations to characterize the system in the classical limit:

\begin{eqnarray}
\dot{x}_{\rm th} &=& -\gamma_{\rm th}[x_{\rm th}+\beta P_{\rm opt}(a)] \label{eq:PT},\\
\ddot{x}_{\rm ac} &=& -\gamma_{\rm ac}\dot{x}_{\rm ac}-\omega_{\rm ac}^{2}x_{\rm ac}+F_{\rm opt}(a)/m_{\rm ac}  \label{eq:mech_mode},\\
\ddot{x}_{\rm lev} &=& 
\begin{cases}
-\gamma_{\rm lev}\dot{x}_{\rm lev}  \,\qquad\qquad\qquad\qquad F_{\rm opt} \leq F_{\rm g},\\
-\gamma_{\rm lev}\dot{x}_{\rm lev}+F_{\rm opt}(a)-F_{\rm g} \,\qquad F_{\rm opt} > F_{\rm g},
\end{cases} \label{eq:lev}\\
\dot{a} &=& -[\kappa/2-i(\Delta+G(x_{\rm th}+x_{\rm ac}+x_{\rm lev}))]a+\sqrt{\kappa_{\rm in}}a_{\rm in}. \label{eq:cav}
\end{eqnarray}

The dynamics of the optical field $a$ given by Eq.~(\ref{eq:cav}) correspond to the typical evolution of the intracavity field under the additional back-action from the photothermal, acoustic, and center-of-mass modes. The field is in the frame rotating at the laser frequency $\omega_{\rm l}$, initially detuned from the cavity resonance frequency $\omega_{\rm opt}$ by $\Delta = \omega_{\rm l}-\omega_{\rm opt}$. The cavity is driven by a field of amplitude $a_{\rm in}$ coupling through the input mirror at a rate $\kappa_{\rm in}$. The coefficient $G=\omega_{\rm opt}/L$, with $L$ being the length of the cavity, is the optomechanical coupling between the mirror and the intracavity field converting a position shift into a change in detuning. The constant $\beta$ is the photothermal response coefficient. The radiation pressure force $F_{\rm opt}$ is related to the optical power within the cavity $P_{\rm opt}$ as $F_{\rm opt} = \hbar G |a|^2 = 2P_{\rm opt}/c$, while $F_{\rm g} = m g$ represents the gravitational weight of the mirror. The quantities $g$ and $c$ respectively indicate the free-fall gravitational acceleration and the speed of light. The mirror's total inertial mass is $m$, while $m_{\rm ac}$ is the effective mass of the acoustic mode of frequency $\omega_{\rm ac}$. The dissipation mechanisms in the system are described by $\kappa$ for the intracavity field, $\gamma_{\rm th}$ for photothermal expansion, $\gamma_{\rm ac}$ for the acoustic mode, and $\gamma_{\rm lev}$ for the center-of-mass motion. We note that there is no direct coupling between the three displacements and that their interaction is enabled only through the optical field.

First we describe the photothermal expansion in Eq.~(\ref{eq:PT}). Even though the mirror is highly reflective, the intensity of the field circulating inside the cavity is high enough that even a small fraction of power absorbed in the coating causes it to expand noticeably. Considering that the beam size near the coating is around \SI{100}{\micro\meter} and that the intracavity power is as high as a few kilowatts, the optical intensity can reach \SI{3}{\mega\watt.\centi\meter^{-2}}, which is even larger than that of LIGO \cite{AbbottObservation2016}. Thanks to the ion beam sputter coating, this intensity is still below the laser damage threshold. Nevertheless, the extremely high optical intensity still causes a local rise in temperature, which leads to expansion and therefore a change in cavity length. We model the photothermal displacement under the empirical assumption of an exponential relation with the intracavity power \cite{marino_canard_2006}, governed by the photothermal coefficient $\beta$ and dissipation rate $\gamma_{\rm th}$. The heating of the mirror results in a decrease in cavity length, indicating a positive value for $\beta$. Note that $\beta$ can be negative if the photothermal heating were to lengthen the cavity \cite{MaPhotothermally2020}. This could happen, for example, if the photorefractive effect were dominant, or if the substrate had a negative expansion coefficient.

The excitation of the acoustic modes is described by Eq.~(\ref{eq:mech_mode}). The steady impact of radiation pressure drives the transverse vibrations of the mirror. Given the high aspect ratio of the substrate disk, these vibrations have a significant impact on the overall dynamics of the cavity. Finite-element analysis of the natural mechanical frequencies of the levitating mirror returns a primary acoustic mode of interest around \SI{30}{\kilo\hertz}. The direct displacement measurement shown in Fig.~\ref{fig:4p5W}(a) gives a natural frequency of 28.(6)~\SI{}{\kilo\hertz} and a damping rate of \SI{30}{\hertz}. The discrepancy between the theoretical and experimental values of mechanical frequency may be due to the imprecision of simulated coating layers and mirror geometry.  Other modes are also observed in the displacement spectrum, but their magnitude is much smaller, and for all practical purposes they can be neglected in the following analysis. We note that the resonant frequencies of the acoustic modes are sensitive to the constraints set by the supporting contacts of the mirror and other imperfections at the time of fabrication.

Finally, in Eq.~(\ref{eq:lev}) we examine the displacement of the center of mass by radiation pressure. This is the degree of freedom linked to optical levitation. When the optical push is weaker than the gravitational weight, the radiation pressure force is fully balanced by the constraint of the mechanical support, and the mirror rests on the stage. Above this threshold the net difference between optical and gravitational forces lifts the mirror to a new equilibrium position, tipping it away from one of the contact points of the supporting structure and modifying the relative detuning of the cavity. The tipping angle is generally very small, with the center of mass displacement being in the nanometre scale as opposed to the size of the mirror of a few millimeters. It is, therefore, reasonable to consider linear forces and displacement (rather than torque and angle) in the equation. Note that the threshold force is not generally equivalent to the gravitational weight of the mirror but higher, $F_{\rm th} \gtrsim F_{\rm g}$, as it is also necessary to account for Van der Waals interaction and other static forces. Also, the damping coefficient $\gamma_{\rm lev}$ is considered to differ from that of the acoustic mode, $\gamma_{\rm ac}$, since the former is mostly affected by air viscosity and not internal friction.

The steady-state solutions of the system dynamics are obtained by setting the derivative terms in Eqs.~(\ref{eq:PT})-(\ref{eq:cav}) to zero:
\begin{align}
& x_{{\rm th}}^{0}=-\alpha|a_{0}|^{2} \label{eq:xth0},\\
& x_{{\rm ac}}^{0}=\frac{\hbar G|a_{0}|^{2}}{m_{{\rm ac}}\omega_{{\rm ac}}^{2}} \label{eq:xd0},\\
& \begin{cases}
x_{{\rm lev}}^{0}=0\;\;\qquad\qquad\qquad F_{{\rm opt}}\leq F_{{\rm mg}}\\
\hbar G|a_{0}|^{2}=mg \;\qquad\qquad F_{{\rm opt}}> F_{{\rm mg}} \label{eq:xl0}
\end{cases},\\
& a_{0}=\frac{\sqrt{\kappa_{{\rm in}}}a_{{\rm in}}}{\kappa/2-i[\Delta+G(x_{{\rm th}}^{0}+x_{{\rm ac}}^{0}+x_{{\rm lev}}^{0})]} \label{eq:a0_NDL},
\end{align}
where $\alpha=\beta\hbar\omega_{c}/\tau_{c}$. It is convenient to define an effective detuning of the cavity: $\Delta_{\rm eff} = \Delta+G(x_{{\rm th}}^{0}+x_{{\rm ac}}^{0}+x_{{\rm lev}}^{0})$. From these equations we can calculate the minimum input power required to optically lift the mirror. Assuming the cavity to be on resonance and static forces on the mirror to be negligible (i.e.\ $F_{\rm th} = F_{\rm g}$), the input power threshold is given by
\begin{eqnarray}
P_{\rm th} = \frac{mg\kappa^2L}{4\kappa_{\rm in}}. \label{eq:Pth}
\end{eqnarray}
When the input power is above threshold, the system reacts by reaching a new equilibrium point where the intracavity power is the same but the effective detuning is different. Thus, above threshold the optical power circulating within the cavity is purely determined by the mass of the mirror and is independent of input power, as suggested by the second line of Eq.~(\ref{eq:xl0}). A similar argument applies when $F_{\rm th} \gtrsim F_{\rm g}$.

The parameter values extrapolated directly from the experiment or inferred from the following analysis are provided in Methods. The threshold input power for levitation is expected to be around \SI{4}{\watt}. In the following sections we will unravel the combined dynamics of the system, starting from an input power well below threshold and proceeding at increasingly higher power until the anticipated	 threshold is exceeded.

\begin{figure}[htb] \centering
\includegraphics[width=0.48\textwidth]{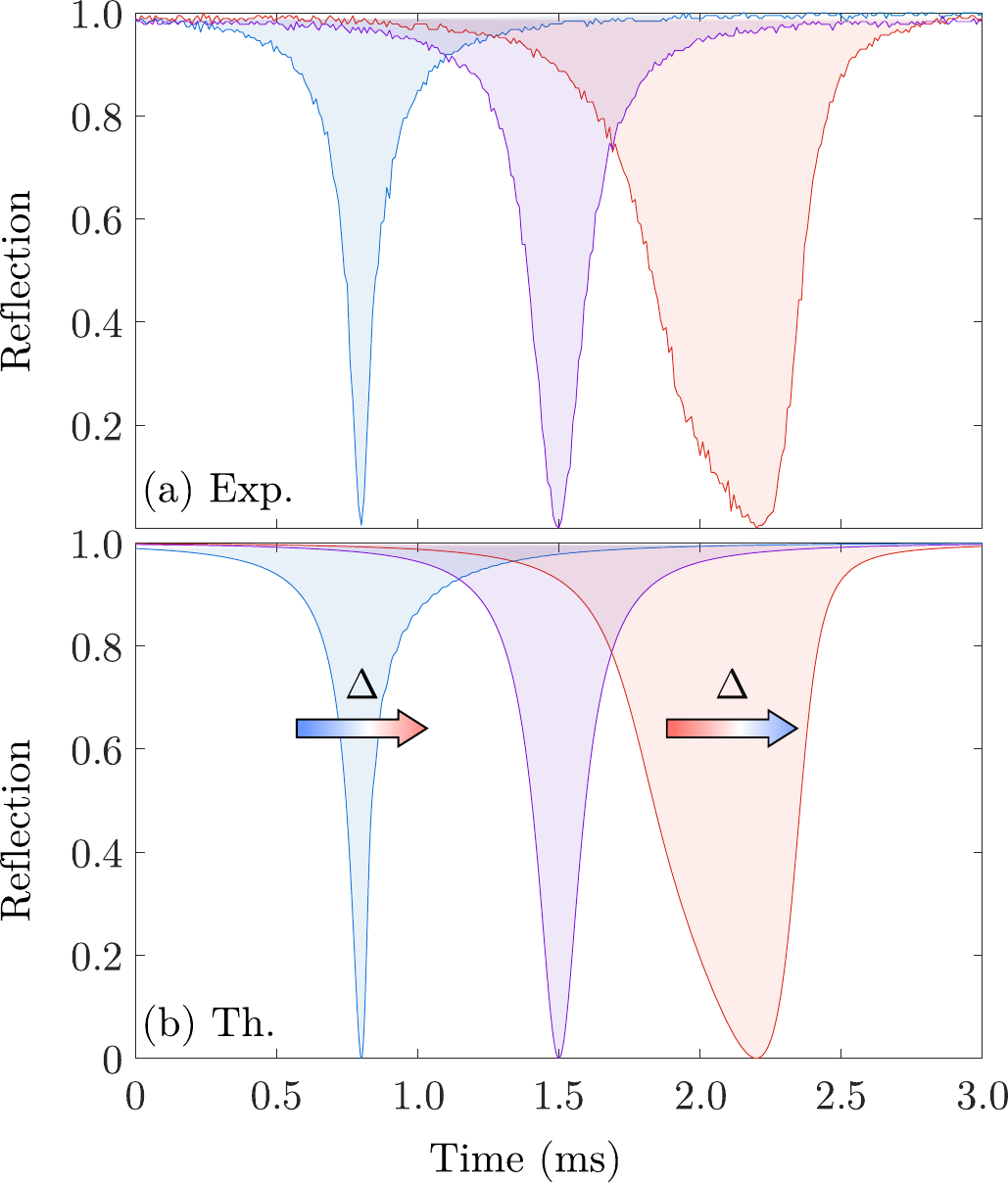} \caption{\textbf{Optical bistability in the cavity}. Exp. and Th. refer to Experiment and Theory. For convenience, we rescale the cavity reflection such that its maximum is normalized to one and its minimum is shifted to zero. The color change of the cavity detuning $\Delta$ indicates the scan direction. (a) Experimental cavity output on reflection at an input power of \SI{180}{\milli\watt} (blue trace for a upward scan and red trace for a downward scan) and \SI{8}{\milli\watt} (purple trace for both scan directions, for reference). The data is obtained by linearly scanning the piezoactuator at a speed of \SI{1.0}{\micro\metre.\second^{-1}} to vary the detuning over time. (b) Numerical solution of Eqs.~(\ref{eq:PT})-(\ref{eq:cav}).}
\label{fig:bistability}
\end{figure}

\noindent\textbf{Optical bistability.} Below threshold the effective detuning at equilibrium is determined only by the photothermal expansion and acoustic modes, $\Delta_{\rm eff}=G(x_{\rm th}^0+x_{\rm ac}^0)$. Equations~(\ref{eq:xth0})-(\ref{eq:xd0}) show that both of these are proportional to the average intracavity power, with thermal expansion being negative and corresponding to a decrease in cavity length while the displacement of the acoustic mode is positive and contributes in the opposite direction. Combining the solutions for these modes into Eq.~(\ref{eq:a0_NDL}) we obtain a cubic equation for the cavity photon number $n=|a_0|^2$. Depending on the system's parameters three possible solutions are possible, with two being stable and one unstable. This bistability phenomenon is well known and has been observed in analogous systems also driven by radiation pressure force \cite{DorselOptical1983, marquardt_dynamical_2006-3} and photothermal expansion~\cite{DorselOptical1983, AnOptical1997}.

In our system, the input laser power is the primary free parameter used to trigger bistability. With sufficiently high power, the two stable solutions can overlap and the system exhibits hysteresis as a function of cavity detuning. Intuitively, we can imagine a scenario where the bottom piezoelectric actuator is set to scan downwards to change the detuning. As the cavity approaches resonance, the power builds up enough thermal expansion in the levitation mirror to compensate for the downward travel of the bottom mirror, resulting in a self-locking response.

In practice, with two effects competing in opposite directions, the course of bistability is determined by the displacement that is most reactive to laser power. We define the ratio of photothermal displacement to acoustic displacement as
\begin{eqnarray}
\zeta= \left|\frac{x_{\rm th}^0}{x_{\rm ac}^0}\right| =\frac{m_{{\rm ac}}\omega_{{\rm ac}}^{2}\beta c}{2},
\end{eqnarray}
for a quantitative evaluation of the dominant process. For $\zeta>1$, the photothermal displacement is dominant over the acoustic displacement, and the cavity resonance shifts towards the blue-detuned regime ($\Delta>0$). For $\zeta<1$ resonance shifts instead towards the red-detuned regime ($\Delta<0$) and bistability is phenomenologically the same as expected in the case of optical lifting. In our experiment we find $\zeta=16$ and photothermal expansion to be dominant, as evidenced by the observation of self-locking when blue-detuning the cavity.

In our system, the appearance of bistability occurs from a minimum input power of about \SI{180}{\milli\watt}, as shown in Fig.~\ref{fig:bistability}. We expose the hysteretic behavior by moving the mirror on the piezoelectric actuator upwards (blue trace, from blue to red detunings) and downwards (red trace, from red to blue detunings). Compared to the typical Lorentzian profile (purple trace, obtained at \SI{8}{\milli\watt}) the resonance appears broadened as the cavity tends to self-lock in the red-detuning regime, and narrowed when encountering the unstable state first by approaching from the opposite direction. This response is easily simulated by numerically solving Eq.~(\ref{eq:cav})-(\ref{eq:PT}) for a detuning varying linearly with time. Thanks to an excellent agreement between the experimental data and theory, we use this data to calibrate the free parameters reported at the start of this section.

\begin{figure*}[htb] \centering
\includegraphics[width=1\textwidth]{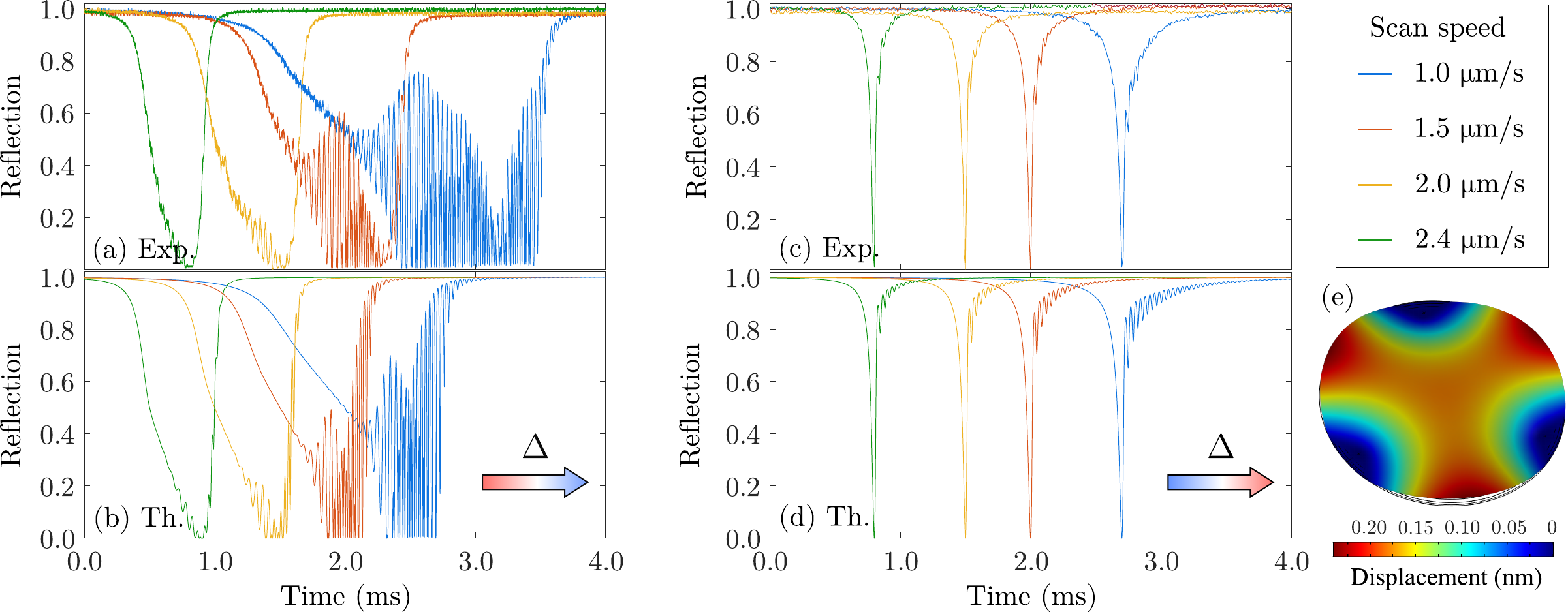} \caption{\textbf{Excitation of the mirror's acoustic mode at different scan speeds of the piezoelectric actuator, at the input power of \SI{500}{\milli\watt}.} Exp. and Th. refer to Experiment and Theory. (a)-(b) Reflection output of the cavity during a downward scan, from red to blue detunings ($\Delta$), showing parametric amplification. (c)-(d) Response of the cavity during an upward scan, from blue to red detunings ($\Delta$). Optical instability prevents the build-up of the oscillations. In both cases the experimental data is shown at the top and the numerical simulation at the bottom. (e) Finite-element analysis of the acoustic mode. The color is proportional to the positive or negative displacement from the rest position.}
\label{fig:PT_acoustic}
\end{figure*}

\noindent\textbf{Excitation of acoustic modes.} Optical bistability becomes more evident at higher input power, when self-locking causes the resonance to become increasingly broad at the same scan speed. At about \SI{500}{\milli\watt} the cavity starts exhibiting a new effect in the form of an oscillatory process, as demonstrated in Fig.~\ref{fig:PT_acoustic}. These oscillations are linked to the excitation of the acoustic modes of the levitation mirror as they get parametrically amplified by the photothermal effect. Similarly to the radiation-pressure induced optical spring effect, the positive photothermal stiffness experienced during self-locking by the system is paralleled by a negative damping coefficient~\cite{BrachmannPhotothermal2016, AltinRobust2017}. The amount by which the natural damping of the acoustic mode is modified by the photothermal interaction can be estimated by the eigenvalues of the Jacobian matrix~\cite{LecamwasamDynamics2020, RoqueNonlinear2020, LiaoTransparency2020, NaumannSteadystate2014, XieInterference2018}.

The excitations are easier to observe when the scan speed is slow enough to enable the full accretion of the oscillations, as seen for example in the \SI{1.0}{}~and~\SI{1.5}{\micro\metre.\second^{-1}} cases in Fig.~\ref{fig:PT_acoustic}(a). Parametric amplification (corresponding to a negative effective damping coefficient) ensues in the red-detuned regime. When the average detuning imposed by the external scan falls on the opposite side of the resonance, the effective damping of the oscillations turns back positive, and the acoustic mode turns quiescent again. Similar oscillations are also excited when scanning in the opposite direction, in Fig.~\ref{fig:PT_acoustic}(b). In this case the red-detuning regime is on the right-hand side of the trace, and since the cavity jumps too quickly to the next stable state, there is not sufficient time for them to develop significantly. A more complete physical picture is given in Supplementary Note 3, where we break down the relative contribution of each degree of freedom during the scan.

We use finite-element analysis to verify that the oscillation frequency (estimated at 28.(6)~\SI{}{\kilo\hertz} by direct measurement), corresponds to a specific vibrational eigenmode of the mirror, shown in Fig.~{\ref{fig:PT_acoustic}}(e). Other vibrational modes were also found. Their participation factor, however, turned out to be at least two orders of magnitude smaller and therefore negligible. This result is in agreement with the full spectrum obtained by the direct measurement of the displacement in Fig.~\ref{fig:4p5W}.

\begin{figure*}[htb] \centering
\includegraphics[width=0.9\textwidth]{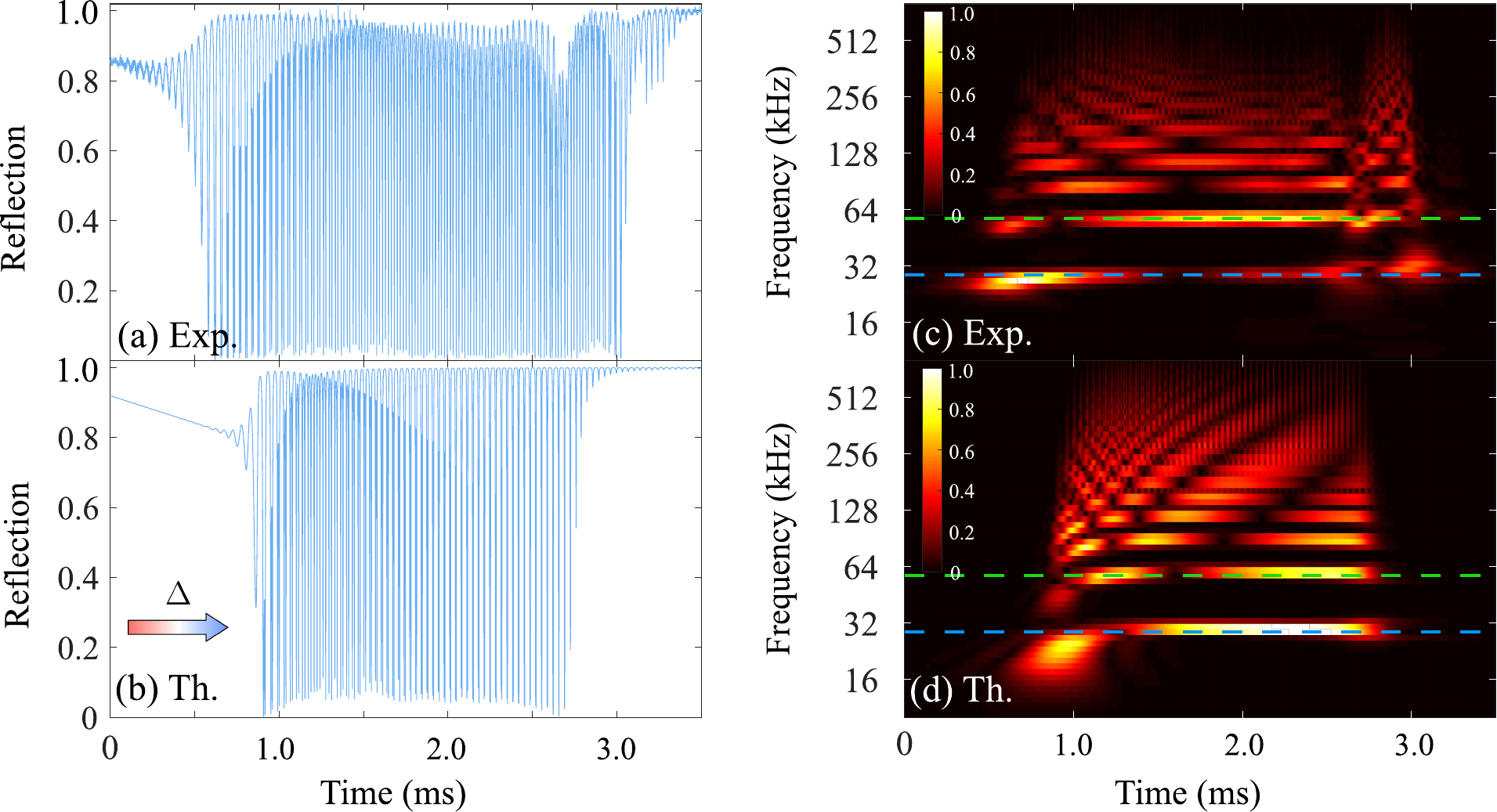} \caption{\textbf{Self-locking cavity response at the input power of \SI{1.9}{\watt}.} Exp. and Th. refer to Experiment and Theory. (a)-(b) These two panels represent the cavity reflection in the time domain, as observed experimentally (a) and simulated numerically (b). The cavity is scanned from red to blue detunings ($\Delta$). (c)-(d) we show the wavelet transform of the corresponding traces. The blue and green dashed lines correspond to the natural acoustic frequency of \SI{28.6}{\kilo\hertz} and its second-order harmonic, respectively.}
\label{fig:acoustic_highPower} 
\end{figure*}

\noindent\textbf{High-order sidebands.} 
The optical cavity field can generate nonlinear amplitude modulations at high intracavity powers. As a result, higher-order stokes and anti-sidebands can be produced in the spectrum of the cavity output field.~\cite{carmon_temporal_2005, carmon_chaotic_opto}.

We tune the input power up to \SI{1.9}{\watt} and present the experimental result in Fig.~\ref{fig:acoustic_highPower}(a). It is shown that a long self-locking envelope carries multi-frequency oscillations. The evolution of the oscillatory frequency can be analyzed by performing the wavelet transform of the time-domain reflection data, as shown in Fig.~\ref{fig:acoustic_highPower}(c). The mechanical oscillation of 28.(6) \SI{}{\kilo\hertz} is excited and parametrically amplified at around \SI{0.2}{\milli\second}. Second-order sidebands start to appear from \SI{0.5}{\milli\second}, and higher-order sidebands are generated when the average detuning approaches the cavity resonance. The blue and green dashed lines refer to the natural frequency of the acoustic mode and its second-order sideband frequency, respectively.

\noindent\textbf{Optical spring.} 
Another feature that we observe in Fig.~\ref{fig:acoustic_highPower} is how the natural frequency of the acoustic mode varies as the mirror interacts with the cavity. This is particularly evident before entering the regime of self-sustained oscillations, where the change in detuning is on average still proportional to the applied linear scan. We attribute this modification to the optical spring effect, which causes a reduction of the effective frequency in the red-detuned regime when the parametric amplification begins, and an increase in the blue-detuned regime when the cavity trails out of resonance~\cite{aspelmeyer_cavity-opto_review}. We note that photothermal effects can also contribute to the optical modification of the natural mechanical frequency~\cite{AltinRobust2017}. The pure optical spring is produced by the back-action of the intracavity optical field via the passive feedback loop $L_1$ between the cavity field and the acoustic mode. The interaction between the photothermal effect and the intracavity field adds an additional feedback loop $L_2$ to the system. The presence of the photothermal displacement works as a feedback path influencing the optical path length of the cavity and then the intracavity field. The variation of the optical field induced by loop $L_2$ further modifies the mechanical susceptibility via loop $L_1$.

The optical spring effect is more prominent in the theoretical plot than in the experimental result, as shown in Fig.~\ref{fig:acoustic_highPower} (c) and (d). This discrepancy might be the result of the following assumptions in our model: firstly, we neglect the direct interaction between acoustic mode and photothermal effects; secondly, we assume that the photothermal displacement is linear to the intracavity power. We can, however, see a fair agreement between the theory and experiment.

\noindent\textbf{Optical lift} Above the threshold input power of \SI{4}{\watt} the mirror should experience optical lift. If static electric forces are sufficiently small, the input power of \SI{4.5}{\watt} used for the plots in Fig.~\ref{fig:4p5W} is expected to satisfy this requirement and to successfully detach the mirror from one of the contact points.

The experiment does not allow direct observation of this phenomenon. Any analysis of the mirror's position is performed through a measurement of the reflective coating. Neither the direct measurement by means of the quadrant detector or the indirect deduction from the cavity response will yield the absolute position of the center of mass, especially taking into account the more consequential dynamics examined so far. As a matter of fact, at this power we do not identify a qualitative difference compared to when the system operates below the lift-off threshold. We remember that, at equilibrium, radiation pressure shifts the detuning point of the cavity so as to achieve the same intracavity power that would be circulating on resonance at the threshold power. Calculating the induced detuning as $|\Delta_{\rm RP}|=(\kappa/2)\sqrt{P_{\rm in}/P_{\rm th}-1}$, we can estimate the centre-of-mass displacement $x_{\rm lev}$ to be on the order of $\Delta_{\rm RP}/G$, i.e.\ about \SI{40}{pm}. For comparison, the distance traveled by the reflective coating due to the acoustic vibrations is proportional to a few linewidths, i.e., \ a multiple of $\kappa/G$, which is equivalent to a few nanometers.

These projections are corroborated by the numerical simulations, which allow us to investigate the photothermal expansion, acoustic vibrations, and mechanical displacement individually. At an input power of \SI{4.5}{\watt} (used in Fig.~\ref{fig:4p5W}), it is expected that the scale of both acoustic and photothermal displacements is on the order of a few nanometres while the mirror is only lifted off the stage by a few picometres. We also compare the simulated system dynamics in the presence and absence of the optical lifting interaction given by Eq.~(\ref{eq:lev}), finding no apparent difference. We conclude that, at this power, the system dynamics are scarcely affected by the optical lift, and it is not possible to identify a visible signature of this effect.

In general, it may be tempting to increase the power well above the threshold to ensure a noticeable detuning, detectable for example, by monitoring the Pound-Drever-Hall signal of the cavity. We note, however, that the scale of the instabilities makes this task hardly implementable, as the oscillations are such as to alter the instantaneous detuning to a much larger extent.

\begin{figure}[htb]
\includegraphics[width=0.49\textwidth]{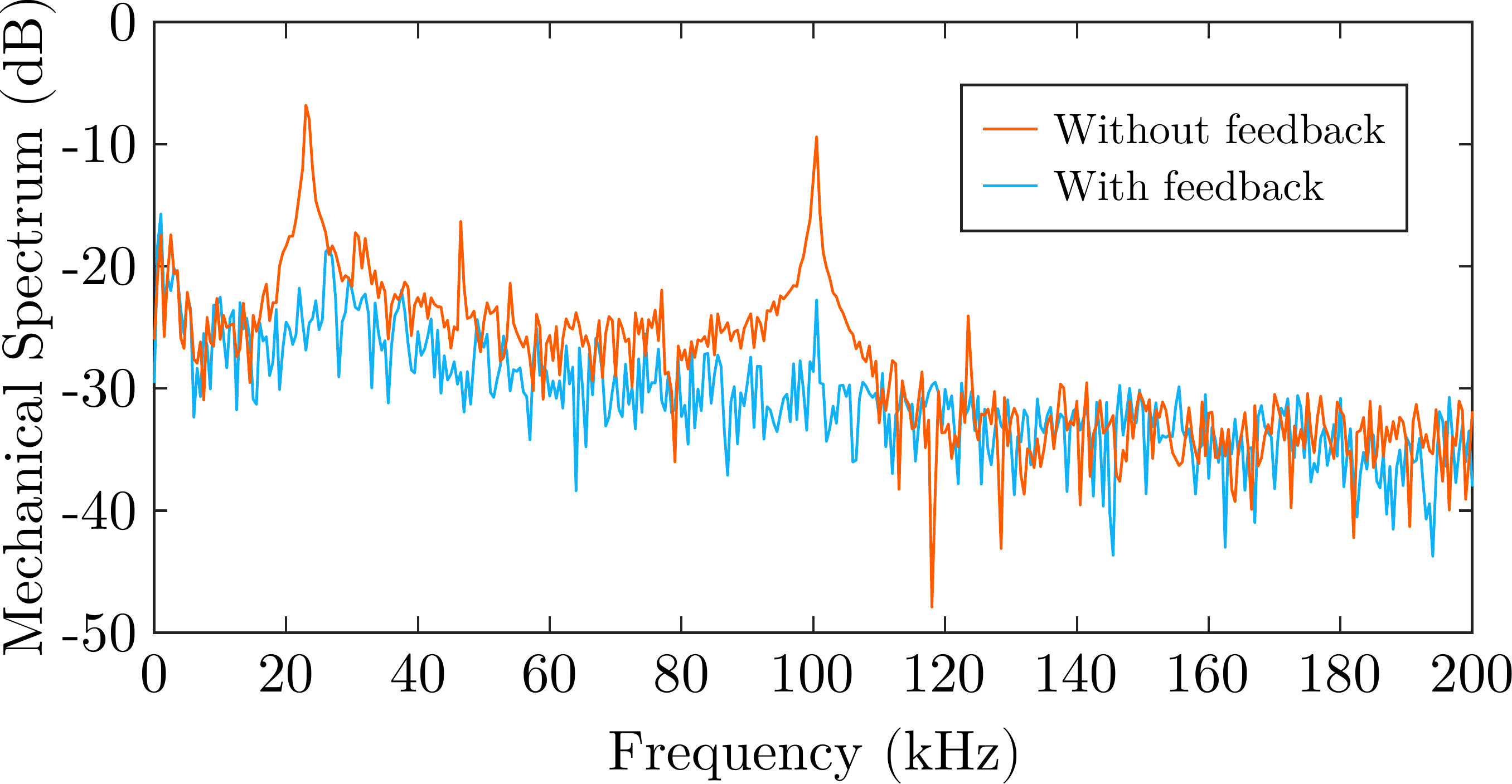} \caption{\textbf{Feedback suppression of the acoustic excitations.} Without feedback, the spectrum of the mechanical displacement shows two natural modes at \SI{30}{kHz} and \SI{100}{kHz} (red trace). With feedback, the two modes are suppressed by more than \SI{10}{\decibel}.}
\label{fig:feedback}
\end{figure}

\noindent\textbf{Feedback cooling of system instability.} The excited eigenmodes of the mirror destabilize the cavity and make it impossible to evince the possible suspension of the mirror on the optical field. Moreover, as the vibrations cover the whole surface of the mirror they would inevitably propagate to the other independent cavities in the final tripod configuration. We apply active feedback~\cite{poggio_feedback_2007, CorbittOpticalDilutionFeedback2007} to the input power in order to suppress such instability and avoid cascading disruptions. We feed the direct displacement measurement of the quadrant detector to an acousto-optic modulator, providing a modulation of the input power and therefore adding an effective drive to the mirror on top of radiation pressure force. Accounting for a delay $\tau$ within the feedback line, the equation for a single acoustic mode is
\begin{eqnarray}
\ddot{x}_{\rm ac} + \gamma_{\rm ac}\dot{x}_{\rm ac} + \omega_{\rm ac}^2x_{\rm ac}&=&[F_{\rm opt}+F_{\rm fb}(t-\tau)]/m_{\rm ac}, \label{eq:feedback}
\end{eqnarray}
with $F_{\rm fb}(t-\tau)=gx(t-\tau)$ being proportional to the total displacement measured by a factor $g$, representing the feedback gain. The delay $\tau$ can be tuned via a phase shifter or a differentiator for optimal results. The effective mechanical susceptibility is obtained by solving Eq.~\ref{eq:feedback} in the frequency domain:
\begin{eqnarray}
\chi_{\rm eff}(\omega)&=&\frac{1}{m_{\rm ac}(\omega_{\rm ac}^{2}-\omega^{2})-g\cos(\omega\tau)+i[m_{\rm ac}\gamma_{\rm ac}\omega+g\sin(\omega\tau)]}. \nonumber\\
\end{eqnarray}
The feedback force modifies the natural frequency and damping rate of the acoustic mode. At $\omega\tau=\pi/2$ the frequency remains unchanged but the damping can be modified to become positive and stabilize the mirror.

To implement feedback control we use a thinner levitation mirror, with a thickness of approximately \SI{30}{\micro\metre} and a mass of \SI{0.966}{\milli\gram}. This mirror displays two major acoustic resonances, one around $\SI{30}{\kilo\hertz}$ and the other at $\SI{100}{\kilo\hertz}$. The system dynamics for this mirror exhibit a further nonlinear phenomenon in the form of a continuous spectrum characteristic of chaotic systems, as we report in Supplementary Note 1. The experimental setup for the feedback control is shown in Fig.~\ref{fig:1_NDL}(d). The modulation of the input power is achieved using an acousto-optic modulator, driven by a \SI{80}{\mega\hertz} harmonic oscillation. The signal detected by the quadrant detector is filtered using a bandpass with its bandwidth from \SI{20}{\kilo\hertz} to \SI{100}{\kilo\hertz}, giving a clean signal containing the information on the mirror displacement. This signal is then phase-shifted,  amplified, and fed to the modulator to vary the amplitude of the optical input field. This process forms a closed feedback loop. We process the mechanical displacement on a spectrum analyzer and disclose the obtained results in Fig.~\ref{fig:feedback}. Tuning the feedback phase accordingly, we can achieve essentially complete suppression of the acoustic excitations (more than \SI{10}{\decibel} for both modes) at powers much lower than the levitation threshold for this particular mirror.

The technique demonstrated is unfortunately not easily extended to different regimes, as it is mostly effective for single-mode applications~ \cite{poggio_feedback_2007}. At high power the instabilities arising from the photothermal interaction prevail and the increasing complexity of the dynamics requires exceedingly stringent feedback parameters. At full operating power for optical levitation, active feedback may altogether be an inadequate choice for stabilization, in which case it may be preferable to consider passive techniques acting directly on the damping properties of the system and therefore addressing the problems closer to the source.

\section*{Discussion}
In this paper, we explored the nonlinear dynamics of a vertical optical cavity where the top reflector consists of a millimeter-scale mirror to be optically levitated. We input up to \SI{4.5}{W} of laser power onto an ion beam sputtered mirror weighing $\approx 1$mg. The spot size of our laser beam is \SI{100}{\micro m} in diameter. This size corresponds to an optical intensity of \SI{3}{MW.cm^{-2}}, which is significantly larger than that of LIGO~\cite{AbbottObservation2016}. We found that the intracavity field bridges the interaction of different degrees of freedom linked to photothermal expansion, acoustic modes, and position of the center of mass. The result is a remarkably complex assortment of dynamics that we proceeded to identify by observing the system's response at different power regimes. We observed optical bistability, parametric amplification, high-order sideband generation, and optical spring corrections to the natural eigenmodes. 

The ideal course of action to reduce the system's instability would be to reduce the optical absorption of the mirror coating significantly. By doing so, radiation pressure would be the major source of interaction between the optical field and the mirror, the combined system would be less elaborate, and any stabilizing effort such as feedback control can be engaged to more specific aspects. Reducing absorption beyond a certain level, however, may be technically challenging. Alternatively, one may attempt to switch the sign of photothermal interaction so that it collaborates with radiation pressure towards both static and dynamic stability~\cite{ballmer_photothermal_2015, AltinRobust2017}. In our system, the photothermal coefficient is $\beta=\SI{8.6}{\pico\metre.\watt^{-1}}$, which is positive. We can flip the sign of this parameter by modifying the thickness of the first coating layer of the levitating mirror~\cite{ballmer_photothermal_2015} or by introducing another photothermal effect with opposite interaction. In the first case, the different thicknesses of the coating would change the mass and the frequency of the acoustic mode of the levitating mirror, as well as the photothermal coefficients. In the other case, adding a new photothermal degree of freedom may require additional components within the optical resonator, modifying the cavity decay rate and finesse together with the effective photothermal response. With this approach, our estimation suggests that the photothermal coefficient can not only be decreased by as little as a factor of two, but also be swapped in sign and changed by as much as two orders of magnitude.

We also expect further explorations of the optomechanical nonlinearity to lead to the observation of interesting stochastic phenomena in the system. Many relevant nonlinear effects such as stochastic resonance have been investigated in optically levitated nanoparticles~\cite{SilerDiffusing2018, RicciOptically2017, FonsecaNonlinear2016, GieselerNonlinear2014}. A milligram-scale mirror driven by an ultra-intense laser opens up a very different parameter regime, where it might be intriguing to extend the concepts and experiments observed at nano-scale towards a new territory and even identify exclusive nonlinear effects.

Despite being focused on a specific optical levitation system, our investigations may offer methods to understand the physics of high-power optomechanical systems.

\section*{Methods}
The top mirror of the levitation cavity presents a high-reflectivity coating of 99.992\% on the curved side of a fused silica spherical cap with a radius of curvature of \SI{25}{\milli\metre}, diameter of \SI{3}{\milli\metre} and thickness of approximately \SI{50}{\micro\metre}. The mirror's coating is obtained by ion beam sputter deposition to reduce loss and absorption. This coating process produces substantial stress on the substrate, as a result it cannot be applied directly to a thin substrate (of few micrometers in thickness) without risk of damage. The coating is therefore finished on a thick substrate with a thickness of \SI{3}{mm}. The mirror is then lapped down to {\SI{50}{\micro\metre}} from the uncoated side. The lapping technique leaves a relatively rough surface on the backside of the mirror, but at least it helps to dissipate the residual stress baked-in from the coating run. The bottom mirror of the cavity is a conventional 1-inch concave mirror, also with high-reflectivity coating on the concave side of the fused silica substrate. This mirror is attached to a piezoelectric actuator to allow scanning of the cavity length, $L_{\rm c}$, and therefore change the detuning. The actuator is pre-loaded via mechanical clamping, improving its stability and performance. The whole cavity is \SI{80}{\milli\metre} long and is enclosed by a monolithic Invar block to reduce thermal fluctuations and isolate the cavity from airflow. A $1050$-\SI{}{\nano\metre} Nd:YAG laser is used to drive the cavity with up to about \SI{15}{\watt} of input power. The mechanical noises produced by the actuator and the cavity mount are too small to be clearly seen in the experimental results presented in this paper.

From our experimental results we infer the parameters of our system to be $L=\SI{80}{\milli\metre}$, $m=\SI{1.116}{\milli\gram}$, $m_{\rm ac}=\SI{0.38}{\milli\gram}$, $\omega_{\rm ac}=2\pi\,\times\,\SI{28.6}{\kilo\hertz}$, $\gamma_{\rm ac}=2\pi\,\times\,\SI{30}{\hertz}$, $\gamma_{\rm th}=2\pi\,\times\,\SI{560}{\hertz}$, $\gamma_{\rm lev}=2\pi\,\times\,\SI{50}{Hz}$, $\kappa=2\pi\,\times\,\SI{730}{\kilo\hertz}$, $\kappa_{\rm in}=2\pi\,\times\,\SI{180}{\kilo\hertz}$, $G=2\pi\,\times\,\SI{3.6}{MHz.nm^{-1}}$, $\beta=\SI{8.6}{\pico\metre.\watt^{-1}}$.

\section*{Data availability} 
The data that support the findings of this study are available from the corresponding author upon reasonable request.



\section*{Acknowledgments}
This research was funded by the Australian Research Council Centre of Excellence CE170100012, Laureate Fellowship FL150100019 and the Australian Government Research Training Program Scholarship.

\section*{Author contributions}
G.G., P.L., G.C. and J.M. conceived the experiment; J.M. and R.L. did the theoretical calculations. J.M. designed and fabricated the device. J.M., J.Q., and G.G. performed the experiment. J.M. wrote the manuscript and all authors contributed to the manuscript. P.L. supervised the work.

\section*{Competing interests}
The authors declare no competing interests.

\section*{Additional information}
Supplementary information is available for this paper.

\end{document}


\title{Supplementary Information for ``Observation of Nonlinear Dynamics in an Optical Levitation System"}

\author{Jinyong Ma, Jiayi Qin, Geoff T.\ Campbell, Giovanni Guccione, Ruvi Lecamwasam,  Ben C.\ Buchler, and Ping Koy Lam$^{}$}\email{Ping.Lam@anu.edu.au}

\affiliation{Centre for Quantum Computation and Communication Technology, Department of Quantum Science, Research School of Physics and Engineering, The Australian National University, Canberra ACT 2601, Australia}

\date{\today}
\begin{abstract}
\end{abstract}

\pacs{}

\maketitle

\subsection*{Supplementary Note 1. Dynamics of a thinner and lighter levitating mirror}
\begin{figure}[htb] \centering
\includegraphics[width=0.48\textwidth]{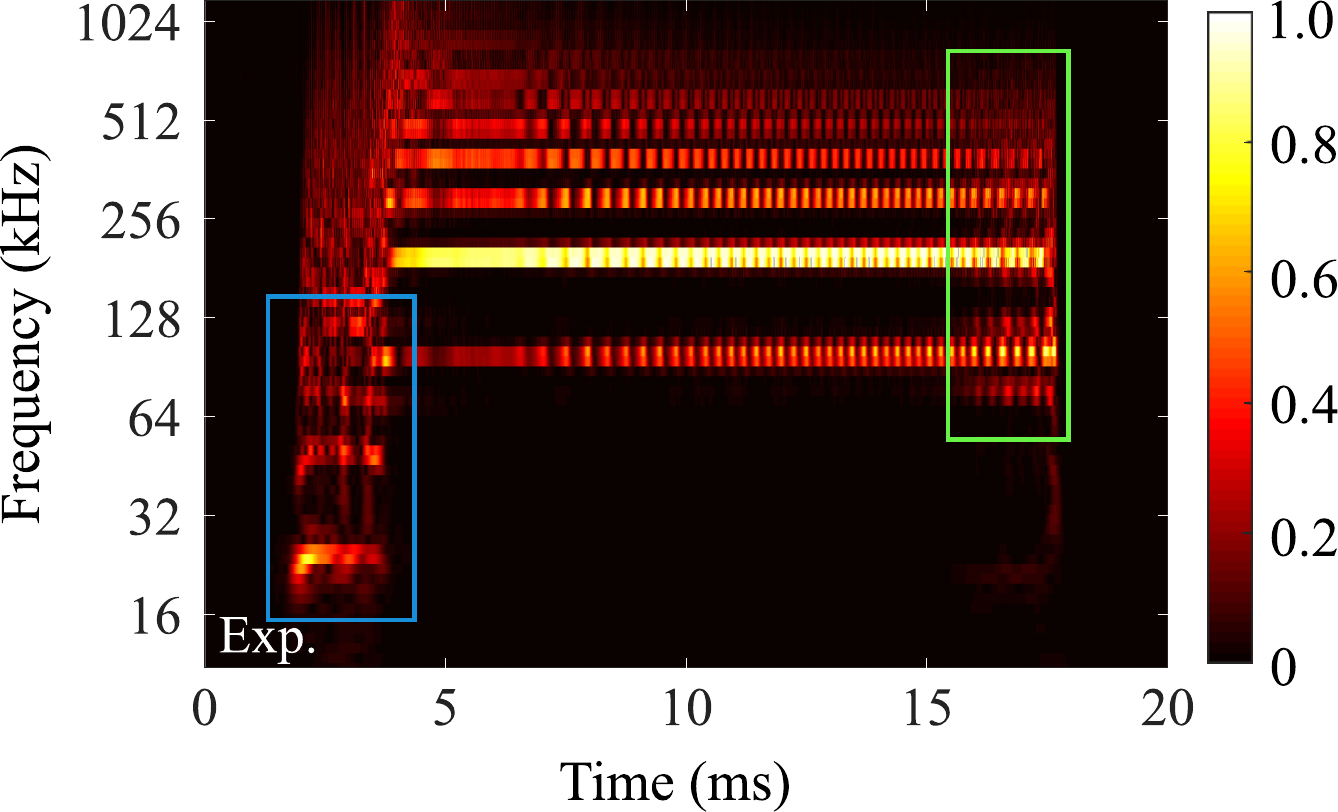}
\captionsetup{labelformat=empty}
\captionsetup{justification=raggedright,singlelinecheck=false}
\caption{Supplementary Figure 1. \textbf{System dynamics for a levitating mirror of  \SI{30}{\micro\meter} of thickness.} We present the wavelet transform of the cavity reflection at the power of \SI{3.3}{\watt} and at the scan speed of \SI{1.6}{\micro\meter/\second}. Unlike the dynamics observed for the 50-\SI{}{\micro\meter} mirror, two acoustic modes of about \SI{20}{\kilo\hertz} and \SI{100}{\kilo\hertz} are excited and present at different scanning displacements. At the start of the scan, the low-frequency mode appears (region enclosed with the blue curve). As the detuning is scanned closer to the cavity resonance, the high-frequency mode is excited while the low-frequency one is suppressed. At the tail of the self-locking train, the two modes are excited simultaneously (region enclosed by the green box).}
\label{fig:30um_wavelete}
\end{figure}

In this supplementary note, we discuss the system dynamics of a thinner levitation mirror. We swap the mirror of \SI{50}{\micro\meter} with one of \SI{30}{\micro\meter} whose mass is $0.966\pm0.003$ \SI{}{\milli\gram}. In terms of Eq.~(9) in the main paper, the power threshold for tipping this mirror is about \SI{3.6}{\watt}. We find that the dynamics of the thinner mirror are more complex. The wavelet transform of the cavity reflection under a downward scan is presented in Supplementary Fig.~\ref{fig:30um_wavelete}. It is shown that two acoustic modes are excited but at two different regimes. At the start of the scan (the region enclosed with a blue box), which is far detuned from the cavity resonance, a low-frequency acoustic mode of about \SI{20}{\kilo\hertz} is present. As the piezoelectric actuator scans closer to the cavity resonance, the low-frequency mode is suppressed while the high-frequency mode of about \SI{100}{\kilo\hertz} is excited instead, as shown in the region enclosed by a blue box. Furthermore, we observe the simultaneous presence of both acoustic modes at the tail of the self-locking train  (the regions enclosed by the green curve), leading to interference between the two high-order frequency components. 

\begin{figure}[htb] \centering
\includegraphics[width=0.9\textwidth]{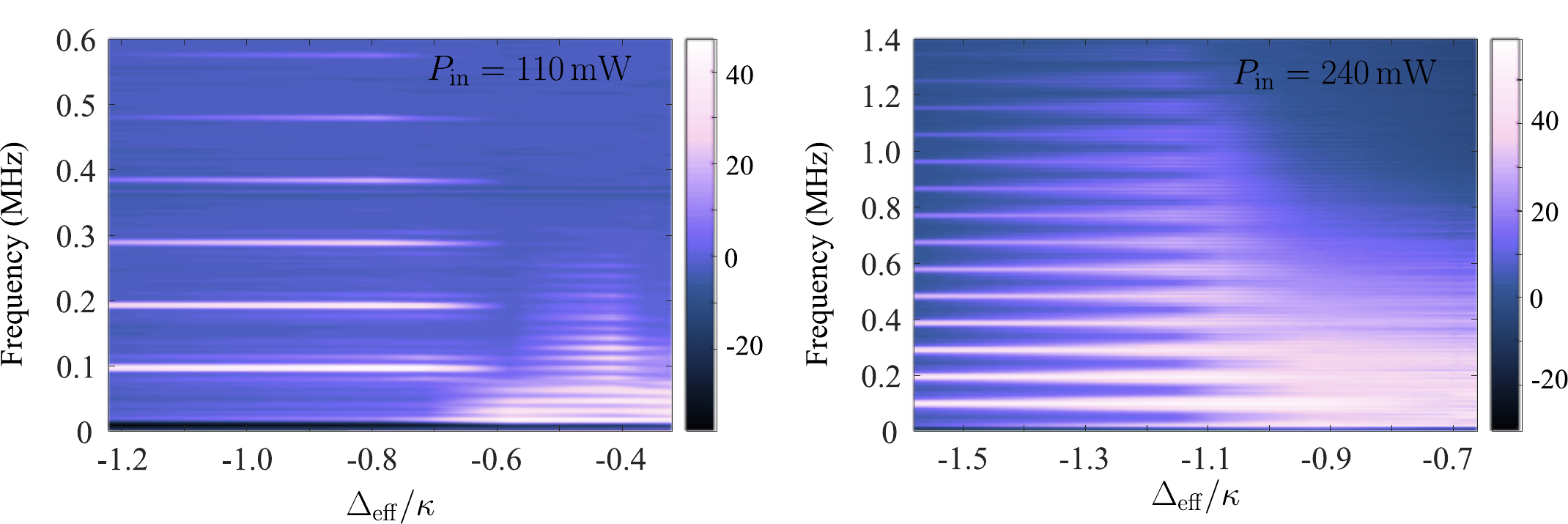}
\captionsetup{labelformat=empty}
\captionsetup{justification=raggedright,singlelinecheck=false}
\caption{Supplementary Figure 2. \textbf{Reflection spectra for the 30-\SI{}{\micro\meter} mirror as a function of effective cavity detuning, under no scan.} (a) Reflection spectra at the power of \SI{110}{\milli\watt}. Agreeing with the scanning case, a low-frequency mode is present at large cavity detunings while a high-frequency mode of about \SI{100}{\kilo\hertz} appears for large cavity detunings. There is a small overlap for the two modes at the detunings around $\Delta_{\rm eff} = -0.6\kappa$. (b) Reflection spectra at the power of \SI{240}{\milli\watt}. Interference of the two excited modes are shown. The spectra become continuous as the detunings is increased.}
\label{fig:30um_NoScan}
\end{figure}

To better understand the interaction between the two excited acoustic modes, we lock the cavity at different red detunings using the photothermally induced self-locking. The spectra of the cavity reflection are measured at each locked detuning, with the results present in Supplementary Fig.~\ref{fig:30um_NoScan}. At the input power of \SI{110}{\milli\watt}, the low-frequency acoustic mode is mainly excited at small cavity detunings while the high-frequency mode is found at large detunings. The higher-order optical sidebands induced by each mode are clearly seen. At a higher power [see Supplementary Fig.~\ref{fig:30um_wavelete}], more high-order frequency components are excited. The peaks for each frequency component tend to merge and become continuous as the detuning is set close to the cavity resonance, which may be an indication of chaos. This effect is primarily due to the strong nonlinear interactions among the photothermal effect, two acoustic modes, and an optical mode.
\subsection*{Supplementary Note 2. Repeatability of experimental results}
\begin{figure*}[htb] \centering
\includegraphics[width=0.9\textwidth]{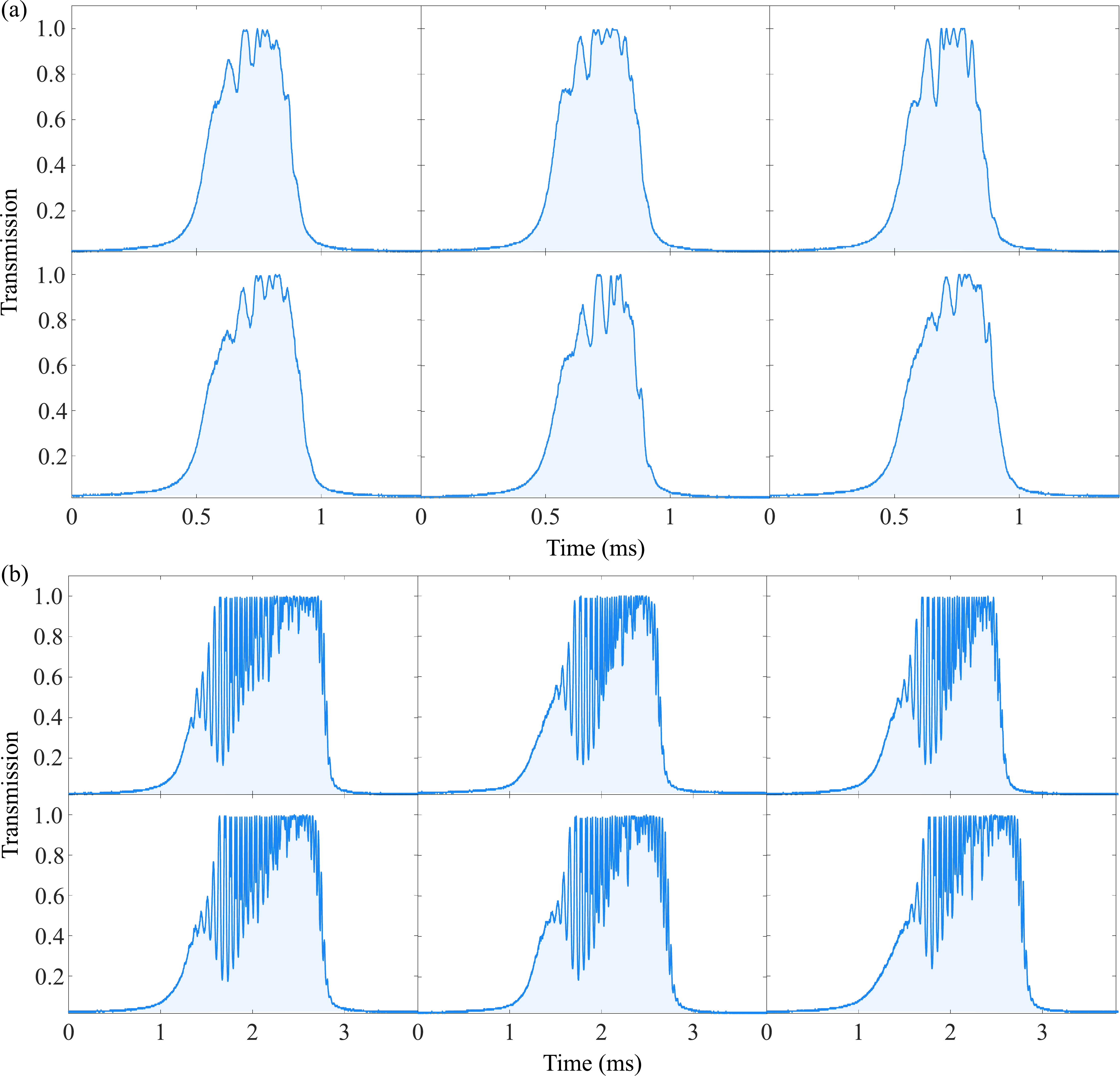} 
\captionsetup{labelformat=empty}
\captionsetup{justification=raggedright,singlelinecheck=false}
\caption{Supplementary Figure 3. \textbf{Multiple measurements of cavity transmission at the input power of \SI{220}{mW} for the 30-\SI{}{\micro\meter} mirror.} The scan speed of the piezoelectric actuator is \SI{1}{\micro m/s} for (a) and \SI{2}{\micro m/s} for (b).}
\label{fig:traces_multi}
\end{figure*}
The experimental data presented in Figs.~2-5 of the main paper are single-time measurements. The results however are highly reproducible, which is demonstrated in Supplementary Fig.~\ref{fig:traces_multi} where we present the result of six different measurements at the same experimental parametric setting for the levitating mirror of \SI{30}{\micro\meter}. The main features of each single measurement remains unchanged though small differences are visible due to the environmental noises from the cavity support and the air flow.

\clearpage
\subsection*{Supplementary Note 3. Relative contributions to the dynamics}
\begin{figure*}[htb] \centering
\includegraphics[width=1\textwidth]{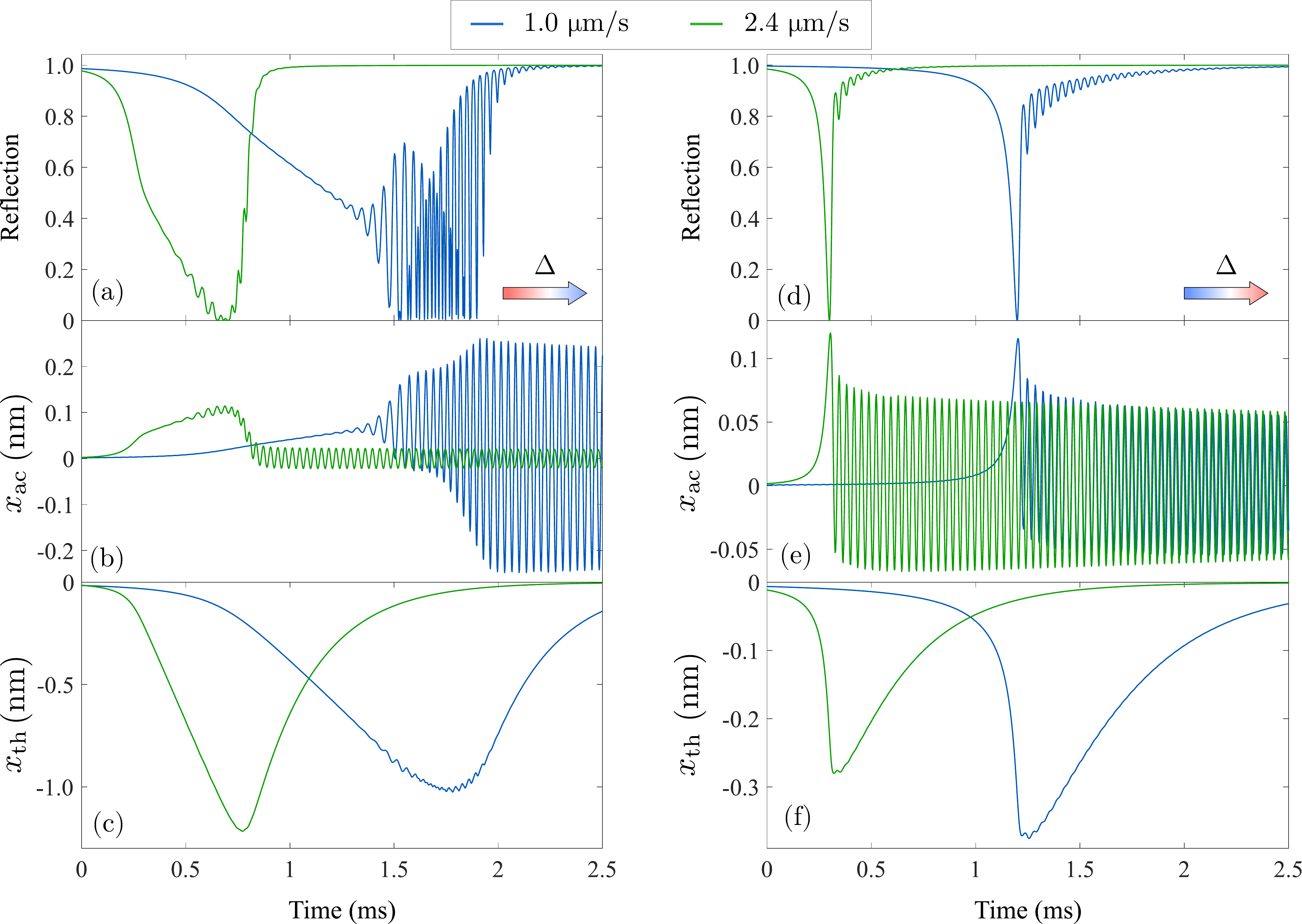} 
\captionsetup{labelformat=empty}
\captionsetup{justification=raggedright,singlelinecheck=false}
\caption{Supplementary Figure 4. \textbf{Time-domain numerical simulations of all variables.} We show the contributions of each degree of freedom to the system dynamics, including cavity reflection, acoustic mode displacement $x_{\rm ac}$, and photothermal displacement $x_{\rm th}$. The green (blue) traces are for the scan speed of \SI{2.4}{\micro m/s} (\SI{1}{\micro m/s}). The parameters used for simulations are the same as the ones given in Fig. 4 of the main paper. (a)-(c) The cavity is scanned downward, from red- to blue- detuned frequencies. (d)-(f) The scan direction is upward, from blue- to red- detuned frequencies.}
\label{fig:traces1}
\end{figure*}

From the comparison of the experimental measurements and the corresponding numerical simulations through Fig. 3-5, it becomes apparent that, depending on the system's conditions, some degrees of freedom influence the dynamics more than others. Here we take the evolution of the system shown in Fig. 4 as an example to show the relative effect of each interaction. In particular, we show in Supplementary Fig. 4 the numerical simulation of all variables in the time domain, giving a visual understanding of how each degree of freedom contributes to the total system dynamics. For clarity we include only the simulation results for two scan speeds (i.e., green traces for \SI{2.4}{\micro m/s} and blue traces for \SI{1}{\micro m/s}). Note that the levitation displacement $x_{\rm lev}$ is not represented because the input optical power is below the levitation threshold, and $x_{\rm lev}$ is identically zero.

We notice that the concurrence of radiation pressure force and photothermal interaction is enough to excite oscillations of the drum mode of the mirror, even if the power is below the threshold for optical levitation. The scale at which this occurs depends on the speed of the scan. Looking at the case when the cavity detuning is scanned from red- to blue-detuned frequencies, it can be seen that the cavity reflection changes dramatically when the scan speed is varied by only a factor of 2.4. The behavior of acoustic mode displacement $x_{\rm ac}$ provides physical insight into this phenomenon. The oscillations of $x_{\rm ac}$ are parametrically amplified as a result of an effective negative damping rate produced by the photothermal effects. At the faster scan speed of \SI{2.4}{\micro m/s} (green trace), the exponential growth of the oscillations is interrupted quite early, and the final amplitude of the oscillations is very small compared to the total displacement due to the photothermal degree of freedom, $x_{\rm th}$. At the slower scan speed of \SI{1}{\micro m/s} (blue trace), the photothermal interaction has more time to drive the oscillations, building up a much larger displacement amplitude for $x_{\rm ac}$. These two different responses are then visibly imprinted on the cavity via the optomechanical interaction. When scanning the cavity detuning from the other direction, there is no such apparent distinction between the two different speeds because the cavity always responds as an impulsive drive. The anti-locking induced by photothermal interaction provides a strong and quick excitation of the acoustic mode in both cases. The swift cavity response disables the manifestation of the dynamics of acoustic mode.